\documentclass[slac_one]{revtex4}
\usepackage{graphicx}
\usepackage{fancyhdr}
\begin{document}

\title{{\small{2005 ALCPG \& ILC Workshops - Snowmass,
U.S.A.}}\\ 
\vspace{12pt} The International Linear Collider beam dumps}

\author{R. Appleby}
\affiliation{The Cockcroft Institute, Daresbury Laboratory, Warrington, WA4 4AD,
England}
\author{L. Keller, T. Markiewicz, A. Seryi, D. Walz}
\affiliation{Stanford Linear Accelerator Center, Menlo Park, CA
94025, USA}
\author{R. Sugahara}
\affiliation{KEK, 1-1 Oho, Tsukuba, Ibaraki 305-0801, Japan}

\begin{abstract}

The ILC beam dumps are a key part of the accelerator design. At Snowmass 2005, the current status of the 
beam dump designs were reviewed, and the options for the overall dump layout considered. This paper 
describes the available dump options for the baseline and the alternatives and considers issues for the 
dumps that require resolution.

\end{abstract}

\maketitle

\thispagestyle{fancy}

\vspace{-0mm}
\section{INTRODUCTION}

The International linear collider (ILC) is a very complicated and expensive project. One aspect of key importance is the main beam dumps, which are required to safely dispose of the high power ILC beams. The work on such dumps was started at the SLC, albeit at much lower power, and continued as part of the TESLA project. However, much more work is needed to obtain satisfactory beam dumps for the ILC. We shall now review the baseline and alternative beam dumps designs.

\vspace{-0mm}
\section{THE ILC DUMP REQUIREMENTS}

The baseline layout of the ILC consists of two linacs, which then branch into two interaction regions. The current choice of crossing angle for the interactions regions are 20mrad and 2mrad. The requirement to dump the main beam after collision then leads us to two full power beam dumps for each interaction region. Furthermore, the need to dump the beam at the end of the linac for commissioning and fast extraction purposes adds two more full power beam dumps to the baseline design. Note that the 500 GeV machine beam power is 11MW and the 1 TeV machine beam power is 18MW; we always consider the dump being rated for the larger power in this paper.

There is also a need to dump the intense beamstrahlung photons generated during the beam-beam interaction at the interaction point. This photon power is around 1MW. This photon beam dump is common for charged beam and photons for the 20mrad interaction region layout, while for 2mrad layout the photon dump is separate.

Hence, there are separate beam dumps rated for full power for all beam lines including tune-up lines, for a total of six beam dumps in the baseline. The tune-up dumps are required to be sufficiently remote from the IP, so that the collider halls can be accessed for detector maintenance while the linac is being tuned, and full beam sent to tune-up dump.

Technically, the elimination of two full power tune-up dumps should be possible, there will be impact on availability which may be partly mitigated by reduced power tune-up dumps (~0.5MW), the cost saving need to be further evaluated, detailed design would need to be made. [1]

\vspace{-0mm}
\section{THE BASELINE TECHNOLOGY - THE WATER DUMP}

The water-based dump [2] is based on the design built in 1967 as the main dump for the SLAC Linac [3], where it is still in use. This dump was designed and built to work at 2MW, but in practice was only used at 800kW. 
The baseline design for the ILC beam dump is based on water vortex dump, rated for 18MW beam. The choice of a water dump for the baseline has many advantages: the water dump has been studied in detail for accelerator projects, the problems of the larger dump design have been noted, and the studies indicate there are no "show-stoppers". The water dump for the TESLA project was studied in detail at DESY [4], with input from several industrial companies.

The basic principle of a water dump is to present the incoming beam with a region of cold, pressurized water. The beam dumps its energy into this water, which rapidly moves away. This presents the next part of the incoming beam with fresh water. The heat is transferred away through heat exchangers. Sufficiently beyond shower maximum the beam is sufficiently large that steel or tungsten plates can be absorb the tail of the beam energy and help reduce the overall length of the dump.  There is also an outside air final cooling stage and many metres of shielding. The water is separated from the vacuum of the extraction line by a thin window - required to be thin enough to avoid the window itself becoming the dump. The window design is key part of the overall dump design.

The water flow velocity is required to be sufficiently high to avoid volume boiling of the water at the tank operating pressure when the dump is accepting the larger spot sizes of the disrupted beam. The dump window is cooled by convection to the water so that its temperature rise during the passage of the bunch train is less than its thermal stress limit.   The spot size of the undisrupted beam must be sufficiently large to prevent window damage.  This will be done through a combination of optical means, an increase in extraction line length after the last optical element, and sweeping the beam across the face of the window.  Beam sweeping can also help prevent volume boiling of the water behind the window; if employed the sweeping mechanism will need to be interlocked to the machine protection system.  The water circuit consists of two closed loops and an external water circuit. The inner water loop is pressurized to 10bar and has a volume of around 18 cubic meters. The length of the dump, including all shielding, is about 25m longitudinally and about 15m transversely.

The control and transport of radioactive byproducts is of central importance to the dump design. Work in ongoing in this area. For example, isotropically produced neutrons contribute to the shielding thickness and careful computation of the neutron fluence is needed. For a deep-tunnel site the forward-peaked muons are stopped in approximately one km of earth.  A shallow-tunnel site may  require a small downward bend of the beam before it enters the dump.  That would necessitate a separate dump for the beamstrahlung followed by the charged beam dump. 

The required R+D items for the baseline are a study of window survivability, and the corresponding computation of radiation damage, measured in displacements per atom (DPA). A window replacement procedure, probably incorporating remote or robotic handling, and schedule can then be developed. A prototype of the window and a beam test are also necessary. The required test beam must give similar energy densities in the window as the full ILC machine. Furthermore, some studies of pressure wave formation maybe necessary.

\vspace{-0mm}
\section{THE ALTERNATIVE TECHNOLOGY - THE GAS DUMP}

The noble gas dump is the alternative design for the ILC beam dump [5,6]. This consists of about 1km of a noble gas (Ar looks the most promising) enclosed in a water cooled iron jacket. The gas core acts as a scattering target, blowing the beam up and distributing the energy into the surrounding iron. Considerable iron is required to successfully transport the heat to the outside water cooling. As in the water dump, the final layer of cooling is an outside air system

This gas dump design may ease some issues such as radiolysis and tritium production, and a gas profile can be exploited to produce a uniform energy deposition along the length of the dump. However, other issues arise such as particle beam heating of the gas and ionization effects. Further studies needed to understand feasibility and benefits of the gas dump. A further possibility is a gas/water hybrid dump, involving the use of a shorter gas dump as a passive beam expander, followed by a small water dump. This option also required further study. A further possibility is for a rotating solid dump immersed in water, or a dump based on some kind of liquid metal.

The required R+D items for the alternative design are studies of gas heating, including ionization effects, and a study of radiation and activation effects. A study of the gas dump windows is also required. A smaller scale prototype of the dump, and some test beam, would also be required.

\vspace{-0mm}
\section{CONCLUSION}

In this paper, we discussed the baseline and alternative designs of the ILC beam dumps. The baseline is a high pressure water dump. Such a dump has been built before for the SLC at much lower power. This is a solid choice for the baseline, although much work is needed. The alternative choice is a gas-based dump. This looks promising, although many further studies and possibly a prototype will be required.

\end{document}